\documentclass[a4paper]{article}
\usepackage{epsfig}

\title{Large-$N$ transitions for generalized Yang--Mills theories in $1+1$ dimensions}

\author{Florian Dubath\\
{\footnotesize florian.dubath@physics.unige.ch}\\
D\'epartement de Physique Th\'eorique, \\Universit\'e de Gen\`eve, \\24 quai Ernest-Ansermet, \\CH-1211 Gen\`eve 4\\}
\date{}

\begin{document}
\maketitle
\abstract{We describe the entire phase structure of a large number of colour generalized Yang--Mills theories in $1+1$ dimensions. This is illustrated by the explicit computation for a quartic plus quadratic model. We show that the Douglas--Kazakov and cut-off transitions are naturally present for generalized Yang--Mills theories separating the phase space into three regions: a dilute one a strongly interacting one and a degenerate one. Each region is separated into sub-phases. For the first two regions the transitions between sub-phases are described by the Jurekiewicz--Zalewski analysis.
The cut-off transition and degenerated phase arise only for a finite number of colours. We present second-order phase transitions between sub-phases of the degenerate phase. }
\newpage
\section{Introduction}
Since 't Hooft's seminal work, the Yang--Mills theory in 1+1 dimensions (YM2) has become a laboratory for testing ideas and concepts about Yang--Mills and also string theory. The YM2 theory has an exact stringy description in the limit of a large number $N$ of colours \cite{C}\cite{GT}\cite{M}. It is also known that one can build generalizations of the YM2 theory \cite{C}\cite{GSY} and that such generalizations have also a stringy behaviour at large $N$ \cite{DLR}. It was also shown that YM2 has different phases, and in particular a third-order transition was present by Douglas and Kazakov \cite{D_K} (here after DK transition).
Recently, new progress has been made in YM2. In particular it was shown that its time evolution could be interpreted as a Brownian motion into the gauge group \cite{AP}\cite{HT}\cite{AAA}. The equivalent of the cut-off transition, well known for Brownian motion, has also been identified in YM2, and is different from the DK one.

The relevant parameter for both cut-off and DK transitions is the area of the  manifold, which plays the role of an inverse temperature. The phase space is then a half-line for the area running from zero to infinity. As fermions do the YM2 state density is limited by 1 and the fermionic picture can be used to help understand the phase structure: at very low temperature the system behaves as a degenerate Fermi liquid. Raising the temperature, we found the cut-off transition and above it a strong interacting phase where the exclusion principle is at work. At high temperature the fermions dilute and finally the density falls down below 1. Above this point (the DK transition) the fermionic nature is irrelevant, and we have a weak interacting system.

Working with generalized Yang--Mills (GYM) theory the phase space opens from the half-line of the YM2 case to a hyper plane. 

All generalized YM2 theories have the same structure. It is therefore possible to capture all essential features of their phase space by studying a particular model. In this paper we pick up a quartic plus quadratic model and describe its phase space and transitions. From this study we deduce the general case. In particular we show that the cut-off and the DK transitions are general features that extend into generalized YM2 and that such transitions coexist with those described by Jurekiewicz and Zalewski \cite{JZ} (here after JZ transitions).

This paper is organized as follows. We first recall in section \ref{S1} how the generalization of YM2 is obtained. We define the model we use in section \ref{quartic}. We present the phase space and our main results in section \ref{Phase}. Detail of the computations are given in the following sections: DK and JZ transition in sections \ref{DKG}, \ref{dilute/strong} and  \ref{JZS}, cut-off transition in section \ref{S3}, transition between the degenerate phases in section \ref{cf_cf}. We draw some conclusions in section \ref{S4}.

\section{\label{S1}Generalized YM2}
The action is the key for building the generalized YM2. Rather than writing the usual action with the $F^{\mu\nu}F_{\mu\nu}$ term, we follow Ref.\cite{C} and use an equivalent action with an auxiliary field $\phi$. For the $d=2$ case, this action is 
\begin{equation}
\mathcal{I}=-\frac{1}{4}\int d^2x\left(i\sum_a\phi_a\epsilon^{\mu\nu}F_{\mu\nu\ a}+\frac{g^2}{2}\sum_a\phi_a\phi_a\right) \ .
\end{equation}
The generalized YM2 theories (GYM2) are obtained by replacing $\frac{g^2}{2}\sum_a\phi_a\phi_a$ by a sum containing other terms of higher order in $\phi$ with other coupling constant. Building a generalized heat kernel equation \cite{GSY} and using the holonomy variable, we obtain a Hamiltonian of the form 
\begin{equation}
H_G=\sum_k\frac{\lambda_kL}{N^{k-1}}C_k\ ,
\end{equation}
with a higher order Casimir operator $C_k$ rather than only the usual quadratic one. This Hamiltonian replaces the YM2 one which is\footnote{Note that the 1/2 is re-absorbed into $\lambda_2$ for the generalized case.}
\begin{equation}
H_2=\frac{\lambda/2L}{N}C_2\ .
\end{equation} 
In the above expressions we have absorbed the coupling constant into  the generalization of the 't Hooft coupling $\lambda_k$, which is held fixed at large $N$ \cite{GSY} 

The YM2 partition function \cite{Mi}\cite{C} on an orientable surface $\mathcal{M}$ of genus $g$, with $p$ boundaries and surface $A$, is a sum over the irreducible representations $R$ of the gauge group :
\begin{equation}
Z_\mathcal{M}=\sum_Rd_R^{2-2g-p}\chi_R(U_1)\dots\chi_R(U_p) e^{-\frac{\lambda_2A}{2N}C_2(R)} \ ,
\end{equation} where $d_R$ is the dimension of the representation and $\chi_R(U_j)$ the character of the holonomy $U_j$. Its generalized counterpart is simply 
\begin{equation}
Z_\mathcal{M}=\sum_Rd_R^{2-2g-p}\chi_R(U_1)\dots\chi_R(U_p) e^{-\sum_k\frac{\lambda_kA}{N^{k-1}}C_k(R)}\ .\label{Zg}
\end{equation}

Until this point the analysis has been completely general. In order to perform the sum over the irreducible representations, we now specify the gauge group. We are interested in $SU(N)$. These groups have irreducible representations labeled by maximal weight $\{ h_i\}$. A Young diagram can be associated to each representation with rows of a length given by $\{h_i\}$. We make the usual change of variables $\{n_i=h_i+\frac{N+1}{2}-i\}$. The computation of the symmetrized quartic Casimir for $SU(N)$ can be found in \cite{DLR}.  We have
\begin{eqnarray}
C_2(\{n_i\})&=&\sum^{N}_{i=1} n_i^2- \frac{1}{N}\left(\sum^{N}_{i=1} n_i\right)^2-\frac{N(N^2-1)}{12}\label{c2}\\
C_4(\{n_i\})&=&\sum^{N}_{i=1} n_i^4-\frac{2N^2-3}{6}\sum^{N}_{i=1} n_i^2
-\frac{4}{N}\sum^{N}_{i=1} n_i^3\sum^{N}_{j=1} n_j\nonumber\\
&&+\frac{6}{N^2}\sum^{N}_{i=1} n_i^2\left(\sum^{N}_{j=1} n_j\right)^2+\frac{N^2-3}{6N}\left(\sum^{N}_{i=1} n_i\right)^2\nonumber\\
&&-\frac{3}{N}\left(\sum^{N}_{i=1} n_i\right)^4
+\frac{N(N^2-1)(11N^2-9)}{720}\label{c4} \ .
\end{eqnarray}

\section{Quartic model \label{quartic}}
We focus on the case of the sphere, i.e. we study the model for a surface $\mathcal{M}$ with $g=0$ and no boundary. For YM2, the partition function reduces to 
\begin{equation}
Z_S=\sum_{n_1>n_2>\dots>n_N}\left(\prod_{i<j}\frac{(n_i-n_j)}{j-i}\right)^2 e^{-\frac{\lambda_2A}{2N}\sum n_i^2}e^{\frac{\lambda_2A(N^2-1)}{12}} \ . \label{pf2}
\end{equation}
As the denominator $\prod_{i<j}(j-i)$ is the same for all the representations, one can see it as a normalization constant and forget it.
Note that the rescaled area $\lambda_2A$ plays the role of an inverse temperature and that in the large-$N$ limit this model is equivalent to fermions in a potential (in the sense that the state density cannot be greater than 1). The transitions we consider arise in the large-$N$ limit for very different values of the rescaled area. 

Rather than dealing with the general case, one can capture the essential features of the GYM by studying the quartic Casimir case\footnote{Models where the higher Casimir is odd do not lead to an energy bounded from below and therefore do not produce a well defined theory.}. We use the case of a GYM2 model with a mix of the quartic and quadratic Casimir instead of only the quadratic one. The Hamiltonian is now given by
\begin{equation}
H_m=\frac{\lambda_2/2L}{N}C_2+\frac{\lambda_4L}{N^3}C_4=\lambda_4L\left(\frac{\mu/2}{N}C_2+\frac{1}{N^3}C_4\right)\ ,
\end{equation}
where $\mu$ is the ratio of the 2 coupling constants $\mu=\lambda_2/\lambda_4$, and we have kept explicit the factor 2 of $\lambda_2/2$ in order to make the contact between the limits $\mu\rightarrow\infty$ and YM2. The partition function is
 \begin{eqnarray}
Z_S&=& \sum_{n_1>\dots>n_N}\prod_{i<j}(n_i-n_j)^2e^{-\lambda_4A\left(\frac{\mu}{2N}C_2(\{n_i\})+\frac{1}{N^3}C_4(\{n_i\})\right)} \ , 
\end{eqnarray}
From now on we will work with the $SU(N)$ group, therefore $C_2(\{n_i\})$ and  $C_4(\{n_i\})$ are given by (\ref{c2}) and (\ref{c4}) and the subsequent substitution.

\section{Phase space \label{Phase}}
The phase space is described by the variables $A\lambda_4$ and $\mu$. We can also recast these two variables into the Jurekiewicz--Zalewski description \cite{JZ} :
\begin{eqnarray}
\beta_2&=&\lambda_2A=\mu\lambda_4A \\
\beta_4&=&\lambda_4A\ .
\end{eqnarray}  
In this paper, we will use both parameterizations.

\begin{figure}[!ht]
 \begin{center}
\includegraphics[width=8cm]{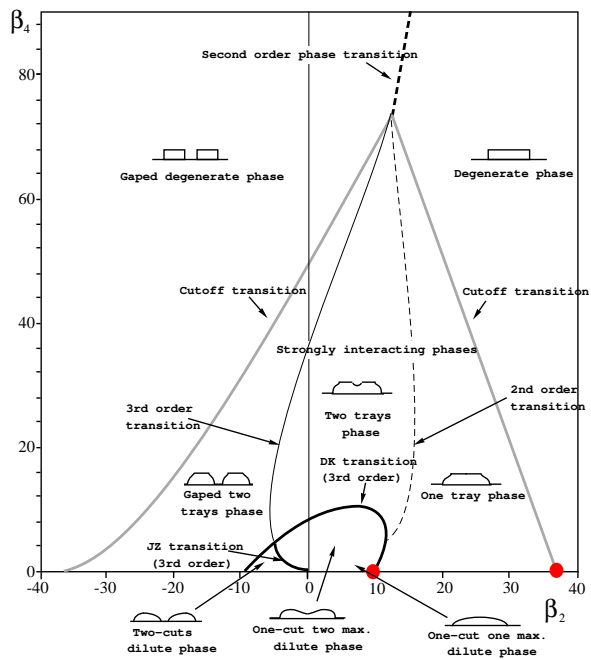} 
\caption{\label{F2}The ($\beta_2$,$\beta_4$) plane for $N=10000$, with the typical form of the density function for each phase. The 2 red dots are the  DK and cut-off transitions for YM2 at the usual values $\left.\beta_2\right\vert_{\rm DK}=\pi^2$ and $\left.\beta_2\right\vert_{\rm cut-off}=4\ln(N)$. The actual location of the thin lines sketched into the strongly interacting phase (phases with tray) are $N$-dependent and not computable with the methods used in this article. The dashed lines correspond to second order phase transitions and (apart for the gray line of the cut-off transition) the continuous lines to third order phase transitions.}
 \end{center}
 \end{figure}

Anticipating our results, we plot the complete phase space (see fig. \ref{F2}) for the $\beta$'s parameterization.

 As in the YM2 case, we have three kinds of phases.  
 
 For small $\beta$'s (high temperature) we have dilute phases where the state density is everywhere below 1. There are two different dilute phases. The first one has a continuous state density and the other one a gapped state density. Between these two phases a third-order phase transition takes place  as expected from JZ work \cite{JZ}. The dilute phases and the transition in between are described in section \ref{DKG}.
 
 Raising $\beta$'s, we cross the DK transition (see sections \ref{DKG}  and \ref{dilute/strong}). The DK  transition is a high-temperature process. It is easier to have intuition about it in the fermionic picture. Raising $\beta$'s  corresponds to lets the temperature go down. The  dilute fermions concentrate until the maximum of the state density is 1. Below this critical temperature, the fermionic nature comes into play. We bring the DK transition to light by reversing this conceptual chain: we compute the state density in a bosonic picture. The DK transition takes place for the value of the $\beta$'s for which the state density goes above 1. 
 
 After the DK transition we enter into strongly interacting phases. Again these are separated in different subspaces, which are the continuation of the JZ ones.
 
 Working at finite $N$, we encounter another phase transition for high $\beta$'s. This is the cut-off transition (see section \ref{S3}). This takes place at low temperature, when the fundamental state ceases to dominate the partition function. This transition was discovered as an analogy of the cut-off transition for random walk on a finite surface \cite{AP}\cite{HT}\cite{AAA}. In the fermionic picture it simply corresponds to the temperature crossing of the Fermi energy. The cut-off transition takes place along the two ($N$-dependent) curves :
\begin{eqnarray}
\beta_2&=&4\ln(N)-\frac{\beta_4}{3}  \label{b_p_1}\\
\beta_2&=&\frac{7}{24}\beta_4-4\ln(N)+\frac{1}{8}\sqrt{\beta_4^2+64\beta_4\ln(N)}\ .\label{b_p_2}
\end{eqnarray} 
These lines draw a triangle with summit 
\begin{equation}
(\beta_2,\beta_4)=\left\{(0,-4\ln(N)),(0,4\ln(N)),\left(\frac{4}{3}\ln(N),8\ln(N)\right)\right\} \ .
\end{equation}
The last summit lies on the line $\mu=\beta_2/\beta_4=1/6$.
 
For larger $\beta$'s (and finite $N$) the system is in degenerate phases. There are two degenerates phases (the fundamental state being given by the trivial or a stepped representation; see section \ref{S3} ), which are separated  by the  $\beta_4=6\beta_2$ line. We show in section \ref{cf_cf} that a second-order phase transition between the two degenerate phases occurs along this line.

\section{DK transition for GYM2\label{DKG}}
The DK transition manifests itself as a saturation of the state density. In the large-$N$ limit, after passing to a continuous variable, one can show that the state density $\rho$ cannot exceed 1. This constraint is not built in the Gaussian matrix model that we use to compute $\rho$ \cite{AAA}; we can therefore track the state density only until its maximum reaches 1.  After which the DK transition takes place. Therefore finding a value of the parameter (the rescaled area) for which $\rho$ reaches the value 1 is sufficient to claim the presence of the DK transition.

We start by presenting the YM2 DK transition. In order to compute the state density, we perform a saddle-point analysis of the partition function (\ref{pf2}), rewriting it as
\begin{eqnarray}
Z_S&=&\sum_{n_1>n_2>\dots>n_N}\prod_{i=1}^{N}e^{N^2S_{\rm eff}} \ ,\\
S_{\rm eff}&=&\frac{2}{N^2}\sum_{j< i}\ln(|n_i-n_j|)-\frac{\lambda_2A}{2N^3}C_2 \label{Seff_2}\ .
\end{eqnarray}
Taking as zero the variation of $S_{\rm eff}$ with respect to $n_i$, and passing to continuous variables through $\frac{n_i}{N}=n$ and $\frac{1}{N}\sum=\int dn\rho(n)$, we obtain for $\rho(n)$ a singular integral equation \cite{MM} 
\begin{equation}
P\!\!\!\!\int dn' \frac{\rho(n' )}{ n-n' } =\frac{\lambda_2A}{2}n\label{SE1}\ ,
\end{equation}
where $P\!\!\!\int$ is the principal value integral. This equation can be solved under the assumption that $n$ varies continuously on a single interval (one-cut solution) and we obtain \cite{MM,NIM} 
\begin{eqnarray}
n&\in&\left[-\frac{2}{\sqrt{\lambda_2A}},\frac{2}{\sqrt{\lambda_2A}}\right]\\
\rho(n)&=&\frac{\lambda_2A}{2\pi}\sqrt{\frac{4}{\lambda_2A}-n^2} \label{rhoYM}\ .
\end{eqnarray}
The maximum of $\rho$ arises for $n=0$ and is equal to 1 if  $\lambda_2A=\pi^2$. 
The DK transition takes place for the value  $\pi^2$ of the rescaled area $\lambda_2A$. 

We tract out the DK transition for the quartic plus quadratic GYM2 following the above steps. 
The effective action that replaces (\ref{Seff_2}) is computed using (\ref{c2}) and (\ref{c4}) and is 
\begin{eqnarray}
S_{\rm eff}&=&-\lambda_4A\left(\frac{1}{N^5}C_4+\frac{\mu}{2N^3}C_2 \right)+\frac{2}{N^2}\sum_{j< i}\ln(|n_i-n_j|)\\
&=&-\lambda_4A\left[\frac{1}{N}\sum_i\left(\frac{n_i}{N}\right)^4
-\left(\frac{3\mu-2}{6}-\frac{1}{2N^2}\right)\frac{1}{N}\sum_i\left(\frac{n_i}{N}\right)^2
\right.\nonumber\\ 
&&-4\frac{1}{N}\sum_i\left(\frac{n_i}{N}\right)^3\frac{1}{N}\sum_j\frac{n_j}{N}+6\frac{1}{N}\sum_i\left(\frac{n_i}{N}\right)^2\left(\frac{1}{N}\sum_j\frac{n_j}{N}\right)^2\nonumber\\ 
&&\left.
-3\left(\frac{1}{N}\sum_i\frac{n_i}{N}\right)^4
+\left(\frac{1-3\mu}{6}-\frac{1}{2N^2}\right)\left(\frac{1}{N}\sum_i\frac{n_i}{N}\right)^2
\right]\nonumber\\&&+\frac{1}{N^2}\sum_{j< i}\left(2\ln(|n_i-n_j|)+\lambda_4A\frac{n_in_j}{3N^2}\right) +{\rm Cst}\ , \label{S_c}
\end{eqnarray}

After taking the variation and changing for continuous variables\footnote{As the continuous limit is valid for large $N$, we can drop the sub-leading contributions.}, we obtain the singular integral equation
\begin{eqnarray}
&&P\!\!\!\!\int dn' \rho(n' )\left(\frac{1}{ n-n' } \right)-\frac{\lambda_4A}{2}\left( 4n^3+\frac{3\mu-2}{3}n\right)\nonumber\\
&&=\frac{\lambda_4A}{2}\left[\int dn' \rho(n' )\left(-12n^2n'-4n'^3+\frac{1-3\mu}{2}n'\right)\right.\nonumber\\ &&\phantom{=}\left. +12\int dn' \rho(n' )\int dn'' \rho(n'' )\left(nn'n''+n'^2n''\right)\right.\nonumber\\ &&\phantom{=}\left.-12\int dn' \rho(n' )\int dn'' \rho(n'' )\int dn''' \rho(n''' )n'n''n'''\right] \ .
 \label{SE2}
\end{eqnarray}
In order to solve this equation for $\rho(n)$ we face two new problems. First if we write the left hand side of this equation into a kernel form we have no longer a Cauchy type integral and, second, we can learn from the study of the cut-off transition (see section \ref{S3}) that the large $\lambda_4A$ distribution of $n$ may have a gap for some range of $\mu$. So we have to be careful where the one-cut hypothesis is valid and work with a two-cut case where it is not. 

Looking at (\ref{S_c}) one can see that symmetric tables (in the sense $n_k=-n_{N-k+1}$) are local minimum of $S_{\rm eff}$. Therefore in the saddle point analysis we will look at distributions $\rho(n)$ which are compatible with this property. That is we restrict ourself to distributions which satisfy  $\rho(n)=\rho(-n)$. This symmetry hypotheses together with the fact that the integral range is symmetric around 0 gives for odd $k$
\begin{equation}
\int dn' \rho(n' )n'^k=0\ ,
\end{equation}
and (\ref{SE2}) reduces to 
\begin{equation}
P\!\!\!\!\int dn' \frac{\rho(n' )}{ n-n' } =\frac{\lambda_4A}{2}\left( 4n^3+\frac{3\mu-2}{3}n\right) \ .          \label{SE3}
\end{equation}
Working with (\ref{SE3}), we have to ensure that the symmetry condition is fulfilled.

\subsection{One-cut solution} 
From the study of the low-temperature case, we expect that, for large enough $\mu$, the distribution $\rho(n)$ will contain no gap. The corresponding solution for (\ref{SE3}) is the so-called one-cut solution.\footnote{One-cut into the plane describing the complexified variable $n$.} Solving (\ref{SE3}) can be done using the same machinery as for (\ref{SE1})  (see \cite{MM}). We obtain the interval $[-a,a]$ given by 
\begin{equation}
a^2=\frac{1}{18}\left(2-3\mu+\sqrt{9\mu^2-12\mu+4+\frac{432}{\lambda_4A}}\right)
\end{equation} 
 and, after some computation, the state density
\begin{equation}
\rho(n)=\frac{\lambda_4A}{2\pi}\left( \frac{3\mu-2}{3}+2a^2+4n^2\right)\sqrt{a^2-n^2}\ .
\end{equation} 
This function fulfills the symmetry condition and has a maximum at $n=0$ or two maxima at \begin{equation}n=\pm\frac{1}{6}\sqrt{4-6\mu+\sqrt{9\mu^2-12\mu+4+432/\lambda_4A}}\ .\end{equation}  
The boundary between these two cases is a curve  into the $(\mu,\lambda_4A)$ plane given by the right branch of 
\begin{equation}
B_1:\lambda_4A=\left(\frac{4}{\mu-2/3}\right)^2\ .
\end{equation} 
Above this curve the maximum arises at $n=0$ and we can compute the value of  $\lambda_4A$ such that $\rho(0)=1$; this gives a curve $T_{\rm DK,1}(\mu)$, which has the asymptotic for $\mu\rightarrow\infty$ ($\beta_4\rightarrow 0$) :
\begin{equation}
T_{DK,1}(\mu)=\frac{\pi^2}{\mu}+\left( \frac{2\pi^2}{3}-4\right)\frac{1}{\mu^2}+\mathcal{O}\left( \frac{1}{\mu^3}\right)
\end{equation}  and cross $B_1$ for $\mu\sim 2.588$.
Below  $B_1$, the two maxima of $\rho$ are equal to 1 for a  curve $T_{\rm DK,2}(\mu)$, which picks a maximum at $\mu=2/3$ and meets $T_{\rm DK,1}$ on $B_1$.

We have to check the one-cut condition. In the two-maxima region, $\rho(0)$ is a minimum and we verify that it is positive. If it is not the case we are no longer in the one-cut case. The equation $\rho(0)=0$ is satisfied on the left branch of  $B_1$ and the one-cut solution breaks down above this curve.

\subsection{Two-cut solution\label{2cuts}} 
\begin{figure}[!ht]
 \begin{center}
\includegraphics[width=8cm]{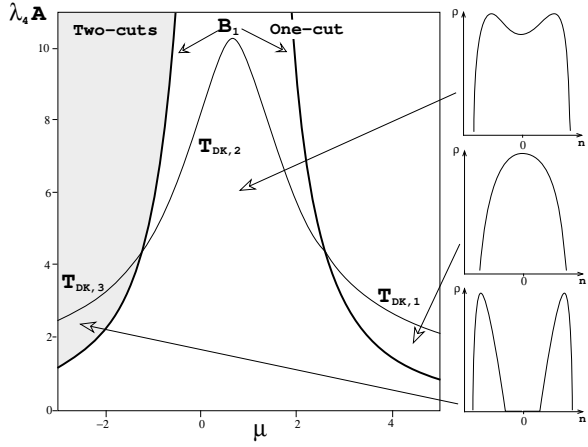} 
\caption{\label{plot1}The ($\mu$,$\lambda_4A$) plane with the one- and two-cut regions and typical form of $\rho(n)$ in each region. The DK transition lines are also plotted.}
 \end{center}
 \end{figure}

For small $\mu$ we have to compute the two-cut solution. Using the symmetry condition and the asymptotic conditions from the general formalism of \cite{MM},  we can compute the state density. This function has support on two segments $[-a,-b]\cup[b,a]$ and is given by \cite{SH}\cite{CMM}
\begin{equation}
\rho(n)=\frac{2\lambda_4A}{\pi}\vert n\vert\sqrt{a^2-n^2}\sqrt{n^2-b^2}\ .
\end{equation}
With $a,b$ satisfying
\begin{eqnarray}
a^2&=&\frac{2-3\mu}{12}+\frac{1}{\sqrt{\lambda_4A}}\ ,\\
b^2&=&\frac{2-3\mu}{12}-\frac{1}{\sqrt{\lambda_4A}}\ ,\\
\end{eqnarray}
$b$ is real above the left branch of the curve $B_1$, exactly where the one-cut solution ceases to be valid. Finding the maximum of $\rho(n)$ we are able to compute the area at which the DK transition takes place. We obtain the curve $T_{\rm DK,3}$ which has the asymptotic for $\mu\rightarrow -\infty$ :
\begin{equation}
T_{\rm DK,3}(\mu)=-\frac{\pi^2}{\mu}-\left( \frac{2\pi^2}{3}+4\right)\frac{1}{\mu^2}+\mathcal{O}\left( \frac{1}{\mu^3}\right)
\end{equation}
and meets  $T_{\rm DK,2}$ when crossing the curve $B_1$. Thus the whole DK transition for $SU(N)$ takes place along a continuous curve. (See fig. \ref{plot1}.)
 
Above the DK transition (for smaller temperature, larger area), the solution $\rho(n)$ cannot be trusted. 

\subsection{Jurekiewicz--Zalewski structure of the dilute phase}
For small coupling values, we are below the DK transition, into dilute phases. We have three regions: the two-cut, the one-cut with two maxima and the one-cut with one maximum, separated by the curve $B_1$; that is, in $\beta$'s language
\begin{eqnarray}
\beta_2&=&\frac{2}{3}\beta_4\pm\frac{1}{4}\beta_a^{3/2}\ .
\end{eqnarray} 
Along the minus branch, between the one- and two-maximum one-cut regions, we have no phase transition. The density $\rho$ and its support $[-a,a]$ are determined by the same functions, which have no singularity of any type along this line; the free energy in the saddle approximation is uniquely determined by $\rho$.

Along the upper branch we expect from the JZ classification \cite{JZ} a third-order phase transition, since a gap opens in the support of $\rho$. That is easily checked by noting that 
\begin{equation}
\left.\rho_{\rm 2-cut}(a,n)\right\vert_{B_1}=\left.\rho_{\rm 1-cut}(a,n)\right\vert_{B_1}=\frac{2\beta_4}{\pi}n^2\sqrt{a^2-n^2}
\end{equation}
and that the support born $a_{\rm 1-cut}$ and $a_{\rm 2-cuts}$ have the same value and first derivative but have a different second derivative. Detailed computation of the dilute phase structure can be found into \cite{AKA}

\section{ Degree of the DK phase transition for GYM2\label{dilute/strong}}
We study the transition between the one-cut one-maximum phase and the corresponding phase above the DK transition. Crossing the DK transition causes the maximum of the state density to be replaced by  a tray. 

Working into the saddle approximation, the free energy depends only on the state density $\rho$. We compute $\rho$ for the one-tray phase. The idea, see  \cite{D_K}, is to set the density at 1 into an interval. We expect a symmetric function and we can parameterize $\rho$ by
\begin{eqnarray}
\rho(n)=0&& \vert n\vert>a\\
\rho(n)=\tilde{\rho}(n)&&c\leq\vert n\vert \leq a \\
\rho(n)=1&&\vert n\vert<c\ .
\end{eqnarray} 
Making the substitution into eq. (\ref{SE3}), we obtain a two-cut problem for $\tilde{\rho}$
 \begin{equation}
P\!\!\!\!\int dn' \frac{\tilde{\rho}(n' )}{ n-n' } =\frac{\lambda_4A}{2}\left( 4n^3+\frac{3\mu-2}{3}n\right) -\ln\left(\frac{n+c}{n-c}\right)\label{SE4} \ .
\end{equation} 
We can use the same setup as in subsection \ref{2cuts} (see \cite{MM} for details): we compute the resolvent for $\tilde{\rho}$
 \begin{equation}
 \omega_0(p)= {\int\!\!\!\!\!\!O}_\mathcal{C} \frac{{\rm d}z}{2\pi i}\frac{\frac{\lambda_4A}{2}\left( 4z^3+\frac{3\mu-2}{3}z\right) -\ln\left(\frac{z+c}{z-c}\right)}{(p-z)}\frac{\sqrt{p^2-c^2}\sqrt{p^2-a^2}}{\sqrt{z^2-c^2}\sqrt{z^2-a^2}}
 \end{equation}
 Deforming the contour to the pole at infinity we also enclose the cut of the logarithm. Taking the discontinuity equation and coming back from $\tilde\rho$ to $\rho$ we eventually obtain the state density 
\begin{equation}
\rho(n)=\frac{\sqrt{a^2-n^2}\sqrt{n^2-c^2}}{\pi}\int_{-c}^{c}\frac{ds}{(n-s)\sqrt{a^2-s^2}\sqrt{c^2-s^2}}
\end{equation}
and from the asymptotic conditions a couple of equations
\begin{eqnarray}
0&=&\lambda_4A\left( a^2+c^2+\frac{3\mu-2}{6}\right)-\frac{2}{a}\mathcal{K}(c/a) \label{ac1}\\
1&=&\lambda_4A\left(\frac{a^2c^2}{2}+\frac{3}{4}(a^4+c^4)+\frac{3\mu-2}{12}(a^2+c^2)\right)-2a\left(\mathcal{K}(c/a)-\mathcal{E}(c/a)\right) \label{ac2}\ ,
\end{eqnarray}
where $\mathcal{K}$ and $\mathcal{E}$ are the standard complete elliptic integrals. 
For a fixed $\lambda_4A$ and $\mu$ ($\beta_4,\beta_2$) we can compute $a$ and $c$ from the above equation. We can also fix other of these four variables.

In order to study the phase transition we look at the $c\rightarrow 0$ limit. Keeping $\mu$ fixed we expand $a$ and $\lambda_4A=\beta_4$ in series of $c$.
The zeroth-order equation matches the values of the dilute phase and the first correction is of order $c^2$. Plugging these solutions into $\rho$ and computing the free energy in the saddle approximation leads to a third-order phase transition. One can also find a special case of this computation into \cite{AKA}
 
\section{Jurekiewicz--Zalewski structure of the strongly interacting phase\label{JZS}}
We present in this section the transition between the different phases with trays (strongly interacting phase). 

From the above section we know the form of the one-tray phase. We compute the two-tray (and no-gap) phase. We are looking for the state density $\rho$  to be an even function of $n$ with a parameterization given by 
 \begin{eqnarray}
\rho(n)=0&& \vert n\vert>a\\
\rho(n)=\tilde{\rho}(n)&&c\leq\vert n\vert \leq a \\
\rho(n)=1&& c>\vert n\vert>d\\
\rho(n)=\tilde{\rho}(n)&&\vert n\vert  \leq d\ .
\end{eqnarray} 
Using  resolvent method we obtain the the state density for the two-tray (no-gap) phase which is given by
\begin{equation}
\rho(n)=\frac{2}{\pi}\int_{-c}^{c}\frac{ds\sqrt{a^2-n^2}\sqrt{c^2-n^2}\sqrt{n^2-d^2}}{(n-s)\sqrt{a^2-s^2}\sqrt{c^2-s^2}\sqrt{d^2-s^2}}\ ,
\end{equation}
we have also four asymptotic conditions. Only two of them are non trivial and are 
\begin{eqnarray}
0&=&\lambda_4A\left( a^2+c^2+\frac{3\mu-2}{6}\right)-2\int_d^c\frac{ds \ s}{\sqrt{a^2-s^2}\sqrt{c^2-s^2}\sqrt{d^2-s^2}} \\
1&=&\lambda_4A\left(\frac{a^2c^2+a^2d^2+c^2d^2}{2}+\frac{3}{4}(a^4+c^4+d^4)+\frac{3\mu-2}{12}(a^2+c^2+d^2)\right)\nonumber \\&&-2\int_d^c\frac{ds \ s^3}{\sqrt{a^2-s^2}\sqrt{c^2-s^2}\sqrt{d^2-s^2}} \ .
\end{eqnarray}
We have two equations for the five variables $\{a,c,d,\lambda_4A,\mu\}$ (which reduce to eqs. (\ref{ac1}) and (\ref{ac2}) in the limit $d\rightarrow 0$).  So we can express $a$ and $c$ as a function of the $\beta$'s and of $d$. For fixed $\beta$'s we obtain a family of solutions parameterized by $d$. All these solutions of the singular integral equation (\ref{SE2}) coexist and, in order to select the one to be used in the saddle-point we need another equation. This can be done by performing the variation of  the effective action with respect to $d$. We can thus obtain the solution $d_m$ as a function of $\beta$'s. Note that without the symmetry condition on $\rho$ the system is under-constrained. Note also that solving the system is a highly non trivial task. 

Finally we have to take into account the fact that, by definition, $b\geq0$ and keep $b={\rm max}(0,d_m)$. So we are in the same situation we will encounter in section \ref{cf_cf} and the phase transition between the one-tray and the two-tray phases  has to be of second order. 

This is consistent with the JZ classification if we focus on $\tilde{\rho}$. The one-tray--two-tray phase transition corresponds to the opening of a new interval into the support of $\tilde{\rho}$ and is thus of order smaller than 3. 

The same can be done for the gaped two-tray phase, looking at a four cuts solution. In this case we have three non trivial asymptotic conditions which together with the minimum of the effective action are enough to fix the position of the trays and gap. According to JZ, the transition between the two-tray and gapped two-tray phases, which is given by the opening of a new gap into the support of $\tilde{\rho}$, has to be of order 3. 

The fact that we can solve the position of trays and gaps is a peculiarity of the quartic plus quadratic model under the symmetry condition. For other models multi-cuts solutions are under-determined and extra constraints have to be imposed \cite{J}.

\section{Cut-off transition for GYM2\label{S3}}
Let us start by recalling the YM2 cut-off transition. We look at the fundamental state. The trivial representation corresponds to the Young diagram with no box. In terms of $\{n_i\}$, this gives
\begin{equation}
R_0:\ \ \ \{n_i\}=\left\{ \frac{N-1}{2},\frac{N-3}{2},\dots,-\frac{N-1}{2} \right\}\ .
\end{equation}
This representation minimizes the Casimir, and thus the energy. As explained in \cite{AAA} the cut-off transition takes place when this representation starts to dominate the other representations in the partition function (\ref{pf2}). This can be estimated by computing the ratio between the partition function contribution from $R_0$ and the one from the "first exited" representation. In the case of $SU(N)$, above the trivial representation, we have the fundamental representation $R_1$ (only one box). We first compute the difference between the Casimir evaluate on this representation and the trivial one :
\begin{eqnarray}
\Delta_2&=&C_2(R_0)-C_2(R_1)=-N\ .
\end{eqnarray}
The ratio between the $R_0$ and $R_1$ contributions is 
\begin{equation}
\frac{Z_S(R_0)}{Z_S(R_1}=\frac{1}{N^2}e^{\frac{-\lambda_2A}{2N}\Delta_2}=\frac{1}{N^2}e^{\frac{\lambda_2 A}{2}}\ ,
\end{equation} and the trivial representation starts to dominate $R_1$ for  the value of $\lambda_2 A$ given by
\begin{equation}
\lambda_2 A=4\ln(N)\ .
\end{equation}
So the system is in a degenerate phase when its rescaled area $\lambda_2 A$ is larger than $4\ln(N)$.

If we want to track the cut-off transition for the GYM2, we have to be careful: we cannot directly extend the computation of the YM2 case, since the state with smallest energy is not necessarily the trivial representation. We make the computation for our quartic plus quadratic model.

\subsection{Fundamental-state candidate}
In order to find the fundamental state, we have to minimize the Hamiltonian over the representations. In fact it is sufficient to find the representation $\{n_i\}$ that gives the smallest value  for the function 
\begin{equation}
E=\left(\frac{\mu}{2N}C_2(\{n_i\})+\frac{1}{N^3}C_4(\{n_i\})\right)\ .\label{E_F}
\end{equation}  
As already explain in sect. \ref{DKG} symmetric table (in the sense $n_i=-n_{(N-i+1)}$) are local minimum of the effective action and also of the function $E$ and, we can guess that the fundamental state is symmetric. For symmetric representations, the $C_4$ has a term proportional to $\sum n_i^4$ minus a term in  $\sum n_i^2$ and the  $C_2$ will modulate this term. Using the fermionic analogy we can see the system as fermions in a Mexican hat potential. For low Fermi energy, we expect to find two sets of fermions, one around each minimum of the potential. Thus we expect a configuration with a gap. Such a gaped configuration corresponds to a Young diagram with a step and we parameterize it as

\begin{eqnarray}
n_i&=&\frac{N+1}{2}-i+q/2{\rm\  ,\   for \ } i\leq N/2 \label{ni1}\\ 
n_i&=&\frac{N+1}{2}-i-q/2 {\  \rm\ ,\    for \ }  i> N/2\label{ni2}\ .
\end{eqnarray} 
Passing to $Q=q/N$, the $Q$-dependent part of the function $E$ takes the value
\begin{equation}
E(R_{Q})=N^2\frac{Q}{8}\left(\frac{Q^3}{2}+Q^2+\frac{3\mu+1}{3}Q +\frac{6\mu-1}{6}\right)+\mathcal{O}(N)\ .
\end{equation}
This function has a minimum, solution of  ${\frac{\partial E(R_Q)}{\partial Q}}=0$, which is given by 
\begin{equation}
Q_m=-\frac{1}{2} +\sqrt{\frac{5}{12}-\mu}\ .\label{Q_m}
\end{equation}
By definition, $Q\geq 0$ and therefore for $\mu\geq\frac{1}{6}$ we have $Q_m=0$, i.e. the fundamental state is the trivial representation. Below $\frac{1}{6}$ the fundamental  state is a step given by the above equation.

\subsection{States near the fundamental-state candidate}
\begin{figure}
\begin{center}
\begin{picture}(274,100)
\multiput(4,30)(75,0){4}{\line(1,0){36}}
\multiput(22,60)(75,0){4}{\line(1,0){22}}
\multiput(44,56)(75,0){4}{\line(0,-1){22}}
\multiput(0,4)(75,0){4}{\line(0,1){22}}
\multiput(22,0)(75,0){4}{\multiput(0,0)(0,3){20}{\line(0,1){1.5}}}
\multiput(0,0)(75,0){4}{\multiput(0,0)(3,0){7}{\line(1,0){1.5}}}
\multiput(3,35)(75,0){4}{$q/2$}
\multiput(24,20)(75,0){4}{$q/2$}

\multiput(44,30)(225,0){2}{\line(0,1){4}\line(-1,0){4}}
\multiput(0,30)(225,0){2}{\line(0,-1){4}\line(1,0){4}}
\multiput(0,0)(75,0){3}{\line(0,1){4}}
\multiput(119,56)(75,0){3}{\line(0,1){4}}
\put(75,30){\line(0,-1){4}\line(1,0){4}}
\put(194,30){\line(0,1){4}\line(-1,0){4}}

\put(48,60){\line(-1,0){4}}
\put(48,60){\line(0,-1){4}}
\put(44,56){\line(1,0){4}}

\put(115,30){\line(0,1){4}}
\put(119,34){\line(-1,0){4}}

\put(150,26){\line(1,0){4}}
\put(154,30){\line(0,-1){4}}

\put(223,0){\line(-1,0){1.5}}
\put(221,0){\line(0,1){4}}
\put(221,4){\line(1,0){4}}

\put(15,67){(A)}
\put(90,67){(B)}
\put(165,67){(C)}
\put(240,67){(D)}

 \end{picture}
 \caption{\label{f2}The four changes for which we check the stability of $R_{Q_m}$.}
 \end{center}
\end{figure}
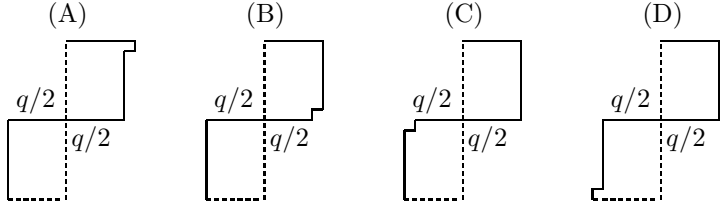

We check  that the step representation is the lowest-energy state.
Looking at the Young diagram, see fig \ref{f2}, we see that there are four ways to change a generic step representation by one box (cases (A)--(D)) .

There is a symmetry between the cases (A) and (D) (resp. (B) and (C)). In fact the variation of the $E$ function and the dimension computation give the same result for the two cases. We obtain
\begin{itemize}
\item $\mu\leq1/6$. For this range of $\mu$, we have $Q_m=0$, only cases (A) and (D) apply. We have
\begin{eqnarray}
\Delta E_{\mu\geq1/6}^{(A)}&=&\Delta E_{\mu\geq1/6}^{(D)}=\frac{\mu}{2}+\frac{1}{6}\ .
\end{eqnarray}
\item $\mu<1/6$ For this range $Q_m$ is given by eq. (\ref{Q_m}) and the four cases are possible :
\begin{eqnarray}
\Delta E_{\mu<1/6}^{(A)}&=&\Delta E_{\mu<1/6}^{(D)}=\frac{5}{24}+\frac{1}{24}\sqrt{15-36\mu}-\frac{\mu}{2}\\
\Delta E_{\mu<1/6}^{(B)}&=&\Delta E_{\mu<1/6}^{(C)}=\frac{5}{24}-\frac{1}{24}\sqrt{15-36\mu}-\frac{\mu}{2}
\end{eqnarray}
and we have  $\Delta E_{\mu<1/6}^{(A)}\geq\Delta E_{\mu<1/6}^{(B)}$ .
\end{itemize}
All these quantities are positive : the step state is the fundamental state.

We now compute the ratio between the dimension of the fundamental state and the one of the near-by cases. We obtain
\begin{eqnarray}
\Delta d^{\rm (A)}&=&\Delta d^{\rm (D)}=\frac{1}{N}\left( \frac{2Q+1}{Q+1}\right)\label{dd1}\ ,\\
\Delta d^{\rm (B)}&=&\Delta d^{\rm (C)}=\frac{1}{N}\left( \frac{2Q+1}{Q}\right)\label{dd2}\ .
\end{eqnarray}
The cut-off transition takes place when the fundamental state dominates all the other states.  The last state to be dominated by  the fundamental state $R_{Q_m}$ is of type (A) or (B), depending on the value of $\mu$.
The ratio between the contribution to the partition function of the  fundamental state and that of the first exited state is of the form
\begin{equation}
\frac{Z_S(R_{Q_m})}{Z_S({\rm Case\ (\star)\ around\ }R_{Q_m}))}=\left(\Delta d\right)^2e^{\lambda_4A\Delta E}\ .
\end{equation} 
The $R_{Q_m}$ representation starts to dominate when the rescaled area is
\begin{equation}
\lambda_4A=\frac{2}{\Delta E}\ln(N)+{\rm Cst} \label{LambdaA}
\end{equation}
where the constant is two times the logarithm of the factor after $1/N$ in (\ref{dd1})--(\ref{dd2}) and can be neglected in the large-$N$ limit. 
Using eq. (\ref{LambdaA}), the cut-off transition takes place :
\begin{itemize}
\item for $\mu\geq1/6$ at
\begin{equation}
\lambda_4A\vert_{\mu\geq1/6}=\frac{4}{\mu+1/3}\ln(N)\ ,
\end{equation}
\item for $\mu<1/6$, using the less energetic case (B),(C), at
\begin{equation}
\lambda_4A\vert_{\mu<1/6}=\frac{4}{-\mu+\frac{5}{12}+\frac{1}{12}\sqrt{15-36\mu}}\ln(N)\ .
\end{equation}
\end{itemize}

Setting $\beta_4=\lambda_4A$ and $\mu=\beta_2/\beta_4$ and solving for $\beta_2$ the two above equations give the phase boundary  (\ref{b_p_1},\ref{b_p_2}).

\section{Phase transition between the two degenerate phases\label{cf_cf}}
In the degenerate phase, the partition function is dominated by the fundamental state. Using the saddle approximation, we will keep only this term in the partition function. We have
\begin{equation}
Z\sim\exp\left(2\ln(d(R_{Q_m }))-\left(\frac{\beta_2}{2N}C_2(R_{Q_m })+\frac{\beta_4}{N^3}C_4(R_{Q_m })\right)\right)\ ,
\end{equation} and the free energy is the exponent divided by $N^2$.
For high $\mu=\beta_2/\beta_4$ we have $Q_m=0$ and the fundamental state is the trivial representation, which has 
\begin{equation}
C_2(R_0)=C_4(R_0)=0 \ \ {\rm and }\ \ d(R_0)=1\ .
\end{equation} 
In this phase the free energy is identically null.
For smaller $\mu$,  $Q_m$ is a function of the ratio $\mu$ and goes to $0$ for $\mu=1/6$. We can compute
\begin{eqnarray}
C_2&=&\frac{1}{4}(Q+1)QN^3\\
C_4&=&\frac{1}{16}(Q^3+2Q^2+2Q+1)QN^5 + \mathcal{O}(N^3)\ .
\end{eqnarray}
As $\lim_{\mu\rightarrow1/6}Q=0$, we have $\lim_{\mu\rightarrow1/6}d(R_Q)=1$; from this and the above Casimir value we see that the free energy is a continuous function.

Using eq. (\ref{Q_m}) we have $\left.\frac{\partial {Q_m}}{\partial\mu}\right\vert_{\mu=1/6}=-1$, so that we can perform the differentiation with respect to $Q$. For the Casimir we get
\begin{eqnarray}
\partial_QC_2&=&\frac{1}{4}(2Q+1)N^3\\
\partial_QC_4&=&\frac{1}{16}(4Q^3+6Q^2+4Q+1)N^5 + \mathcal{O}(N^3)
\end{eqnarray}
and for the dimension using the parameterization (\ref{ni1}) and- (ref{ni2}) we have
\begin{eqnarray}
\left.\partial_Q\ln(d(R_Q))\right\vert_{Q=0}&=&\left.\partial_Q\left(\sum_{i=1}^{N/2}\sum_{j=N/2+1}^{N}\ln(j-i+2Q)\right)\right\vert_{Q=0}\nonumber\\
&=&\sum_{i=1}^{N/2}\sum_{j=1}^{N/2}\frac{2}{N/2+j-i}\nonumber\\
&=&\sum_{\ell=-N/2+1}^{N/2-1}\frac{2(N/2-\vert \ell \vert)}{N/2+\ell} \ .
\end{eqnarray}
Making use of the digamma function $\Psi$ and of its asymptotic behaviour $$\lim_{x\rightarrow\infty}\left(\psi(x)-\ln(x)\right)=0$$ we obtain, for large $N$, 
\begin{equation}
\left.\partial_Q\ln(d(R_Q))\right\vert_{Q=0}\sim N\left(\frac{1}{2}-\ln(2)\right)\ .
\end{equation}
Collecting all the term we get for the derivative of the free energy
\begin{eqnarray}
\left.\partial_\mu F\right\vert_{\mu=1/6}&=&\left(\frac{\beta_2}{8}+\frac{\beta_4}{16}\right)+\mathcal{O}(1/N) \ ,
\end{eqnarray}
and the derivative of the free energy is not continuous across the $\beta_4=6\beta_2$ transition line.
The system undergoes a second-order phase transition along this line.

\section{Concluding remarks \label{S4}}  
\begin{figure}[!ht]
 \begin{center}
\includegraphics[width=8cm]{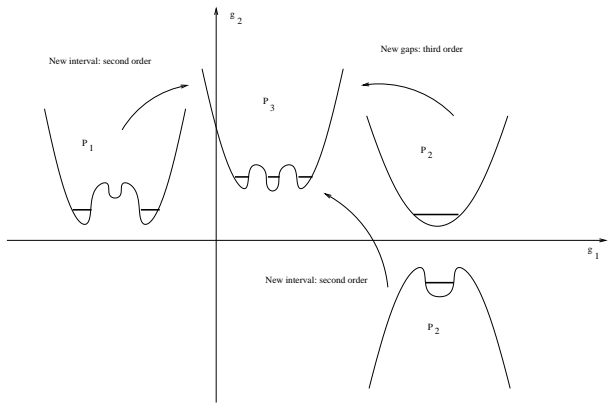} 
\caption{\label{six}Cartoon of the phase structure of the sixth order model with quadratic term fixed. We have used notations of \cite{J}. Note that the lower part of the P$_2$ phase (with $g_2<0$) is unstable.}
 \end{center}
 \end{figure}

The fermionic analogy leads to the conclusion that the three kinds of phase(degenerate, strongly interacting, and dilute) and the DK and cut-off transitions are completely general features that we will find in any GYM2 model.  
Keeping in mind the fact that the higher Casimir of any model has to be of the form $C_{2m}$ in order to guarantee a Hamiltonian bounded from below, the state density will have as a support the union of $k$ intervals with $k\in [ 1,m-1]$. The number of intervals will be a function of $\{\beta_j\}=\{\lambda_jA\}$.  Using the fermionic analogy with fermions in a potential given by a polynomial of degree $2m$, we find intuitively that each local maximum can rise above the Fermi level. 
There exists a region of phase space where the state-density support is only one interval. Moving away from this region gaps will open, splitting the unique interval in different pieces. In the dilute phase this structure is equivalent to the one described in \cite{JZ} from which we can deduce the order of the phase transitions between the different dilute phases. 
As an illustration we can consider the case of the 6th-order model developed in \cite{J}. in this model we have a fixed quadratic term plus a quartic term (with its coupling $g_1$) and a 6th-order term (with its coupling $g_2$). We have then 4 kind of potential for our fermions as shown in figure \ref{six}.
 Looking at the state-density support we deduce the order of the phases transitions according to \cite{JZ}. The precise location of the phase boundary can be found in  \cite{J} and the phase structure is in aggrement with the one we have sketched. 

This picture is also valid in the case in the strongly interacting phase, but each interval of the density of state show a tray. Again argument of \cite{JZ} can be used to obtain the order of the phase transitions between the different tray-phases. 

 In the degenerate phase the state-density is a collection of pieces with value 1 separated by gaps. The state-density support structure depends only on the ratio of the different $\beta$'s and  transitions between this different sub-phases have to be of second order.

\subsection{Remark about $U(N)$}
We have worked with $SU(N)$ groups and want to ask now : What about $U(N)$ ones? The main difference between the $U(N)$ and $SU(N)$ representations is the fact that $SU(N)$ are invariant under translation, i.e. the representation $\{ h_i\}$ is equivalent to $\{ h_i+\ell\}$ with $\ell$ an integer. 

$U(N)$ can be split into $SU(N)\times U(1)$ in the language of the $\{n_i \}$'s; the $U(1)$ part is the "center of mass" position $y=\frac{1}{N}\sum n_i$. The $SU(N)$ is given by the $m_i=n_i-y$, then the sum of the $m_i$ is null. 
For the YM2 case, the quadratic Casimir for $U(N)$ is given (up to some constant term) by 
\begin{eqnarray}
C_2(U(N))&=&\sum (m_i+y)^2=\sum m_i^2+2y\sum m_i+Ny^2\nonumber\\
&=&\sum m_i^2+Ny^2=C_2(SU(N))+Ny^2 \ .
\end{eqnarray}
The $y$ is decoupled from the $SU(N)$ part. Performing the sum over $y$ into the partition function gives an overall normalization and all the considerations on the $SU(N)$ case apply to the $U(N)$ one. 
For GYM2, things are different. Higher Casimirs $C_k$ ($k >2$), couple the $SU(N)$ and $U(1)$ parts. The principal term of $C_k$ for $U(N)$ is
\begin{eqnarray}
\sum n_i^k=\sum(m_i+y)^k=\sum m_i^k +\sum_{j=1}^{k}\pmatrix{j\cr k}y^j\sum_i m_i^{(k-j)}\ ,
\end{eqnarray}
that is the term belonging to the $SU(N)$ Casimir of order $k$ and $y$ dependant terms.
For a chosen GYM2 model (with higher Casimir of even order) and with ratio between the higher Casimir coupling and the other ones $\mu_j$. We collect all the $y$ terms and obtain a polynomial of degree $k$ in $y$.  Unlike $k=2$ case, polynomial coefficients are given by sum of $SU(N)$ Casimirs of order smaller than $k$ and depend on $\mu$'s. 
Plugging this expression into the $U(N)$ partition function and performing the sum over $y$ (which is possible until $k$ is even)  we obtain a function of the $C_j^{SU(N)},\ j<k$ which multiply the $SU(N)$  part. This term can be re-absorbed into an effective action which now depend on all the $SU(N)$ Casimirs of degree $\leq k$. However the $U(N)$ case can be analyzed using $SU(N)$ one with Hamiltonian containing product and power of Casimirs.

\section*{Acknowledgments}
I want to thanks L. Alvarez-Gaum\'e for his patience and his availability, M. Blau for let me know ref \cite{AKA}. M. Mari\~no, M. Weiss and M. Ruiz-Altaba for useful discussions, the CERN th-unit for kindly hosting me during the redaction of this work. Alice and little Fabio for let me sleep enough to be able to work...
This work was partially support by the Swiss national fund (FNS).

\end{document}